\documentclass[conference]{IEEEtran}
\IEEEoverridecommandlockouts
\usepackage{cite}
\usepackage{amsmath,amssymb,amsfonts}
\usepackage[english]{babel}
\usepackage[utf8]{inputenc}
\usepackage[noend]{algpseudocode}
\usepackage{graphicx}
\usepackage{textcomp}
\usepackage{subcaption}
\usepackage{xcolor}
\usepackage{bm}
\def\BibTeX{{\rm B\kern-.05em{\sc i\kern-.025em b}\kern-.08em
    T\kern-.1667em\lower.7ex\hbox{E}\kern-.125emX}}
    
\begin{document}

\title{Predictive and Proactive Power Allocation For Energy Efficiency in Dynamic OWC Networks\\
}

\author{\IEEEauthorblockN{ Walter Zibusiso Ncube, Ahmad Adnan Qidan, Taisir El-Gorashi and Jaafar M. H. Elmirghani}
}
\maketitle
\begin{abstract}
Driven by the exponential growth in data traffic and the limitations of Radio Frequency (RF) networks, Optical Wireless Communication (OWC) has emerged as a promising solution for high data rate communication. However, the inherently dynamic nature of OWC environments resulting from user mobility, and time-varying user demands poses significant challenges for enhanced and sustainable performance. Energy efficiency (EE) is a critical metric for the sustainable operation of next generation wireless networks. Achieving high EE in dynamic OWC environments, especially under time-varying user distributions and heterogeneous service requirements, remains a complex task. In this work, we  formulate a discrete-time EE optimisation problem in a dynamic OWC to maximise energy efficiency through determining  user connectivity and power allocation. Solving this problem in real time is computationally demanding due to the coupling of user association and power allocation variables over discrete time slots. Therefore, we propose a Probabilistic Demand Prediction and Optimised Power Allocation (PDP-OPA) framework which predicts user arrivals, departures, and traffic demands. Based on these predictions, the framework proactively determines AP-user associations and allocates power dynamically using a Lagrangian-based optimisation strategy. Simulation results demonstrate that the proposed PDP-OPA significantly enhances system performance, providing solutions close to the optimal ones. The proposed framework improves energy efficiency, sum rate, and bit error rate (BER) compared to distance-based user association and uniform power allocation, validating its effectiveness for real time and adaptive resource management in dynamic OWC systems.

\end{abstract}


\section{Introduction}
\IEEEPARstart{O}ver the years, the demand for data connectivity has seen exponential growth that shows no signs of slowing down \cite{9144301, 8636954, 8792135, 9040264}. As society becomes increasingly dependent on information and communication technologies (ICT), the need for efficient, and reliable networks has become crucial. Additionally, the radio frequency (RF) spectrum, which has been the backbone of wireless communication, is now congested and unable to cope with the rising demand \cite{9064520}. Alternative solutions are being sought to alleviate this congestion and ensure seamless connectivity and Optical Wireless Communication (OWC) is viable contender. OWC systems utilise the visible light spectrum, through Light Emitting Diodes (LEDs), or infrared radiation, through Laser Diodes such as Vertical-Cavity Surface-Emitting Lasers (VCSELs), for high-speed data transmission \cite{7482669, 5711640}, offering several advantages over RF-based systems \cite{6852089}. These advantages include significantly higher bandwidth, immunity to electromagnetic interference, and improved security \cite{OWC_survey2, 10053990}.

OWC is also inherently energy efficient due to its ability to transmit data over high directional and confined optical signals, which reduces power consumption. This intrinsic energy efficiency has been further enhanced by recent advancements in network design and transmission schemes as in \cite{SE_in_IMDD_Haas,QAM-OFDM2, QAM-OFDM, ZeroFosing1, Energy_Efficient_Resource_Allocation_for_Mixed_RF/VLC_Heterogeneous_Wireless_Networks, Mingqing_EE }.
In \cite{SE_in_IMDD_Haas}, a unipolar orthogonal frequency-division multiplexing (OFDM) modulation scheme was proposed for OWC to enable higher data rates with lower power consumption, making it suitable for energy-efficient  Intensity Modulation/Direct Detection (IM/DD) systems.  In \cite{QAM-OFDM2}, an OWC transceiver was designed and   quadrature amplitude modulation (QAM) in conjunction with OFDM and orthogonal space–time block coding  was employed for enhancing the performance of OWC networks, achieving  high data rates in a range of gigabits per second  (Gbps) over a communication distance of 200 meters, while maintaining low complexity and enhanced energy efficiency, even under realistic channel conditions. Similarly, in \cite{QAM-OFDM}, a high-performance laser diode (LD) system was designed, where QAM-OFDM  was used to achieve high data rates with very low bit error
rate  (BER) over various communication distances. In \cite{ZeroFosing1}, zero forcing (ZF) was employed in a multi-cell OWC network for interference management, resulting in significantly improved energy efficiency  compared to other conventional interference management approaches, making it a suitable transmission scheme for large-scale indoor communication networks. In \cite{Energy_Efficient_Resource_Allocation_for_Mixed_RF/VLC_Heterogeneous_Wireless_Networks},
resource allocation strategies were introduced for maximising energy efficiency in hybrid RF-OWC networks, highlighting improved energy performance compared to stand alone RF and OWC systems. Additionally, in \cite{Mingqing_EE},  the energy efficiency of a VCSEL-based OWC system was eventuated and compared to terahertz (THz) systems, where OWC has been shown to be more energy-efficient, even without performance optimisation.

{In multiple-input and multiple-output (MIMO) systems, power and resource allocation has been widely studied in RF and optical wireless communication systems, to name a few \cite{Pwr5, Pwr2, Pwr6}. In \cite{Pwr5}, a joint user association and power allocation problem was solved using dual projected gradient and successive convex approximation algorithms, achieving better performance compared to  traditional methods by incorporating fairness, load balancing, and power control. A framework  was introduced in \cite{Pwr2} to minimise total power consumption by jointly optimising transmit power and the number of active access points while meeting user spectral efficiency targets. From a quality of service (QoS) perspective, an adaptive power allocation algorithm was introduced in \cite{Pwr6}, where users were classified by QoS demands, and power distribution was adjusted accordingly, improving efficiency under varying conditions.}

{Practical OWC deployments are highly dynamic, constantly subject to time-varying user distributions and requirements. Managing user mobility is crucial in such dynamic environments to ensure seamless connectivity, minimise handover disruptions, and optimise resource allocation, and consequently, enhance energy efficiency by reducing redundant transmissions and maintaining stable network performance \cite{mobility1,RWP_Paper,mobility+classes,LB}. To manage user mobility, a user tracking and mobility management framework  was proposed in \cite{mobility1}  to ensure continuous connectivity for mobile users. The approach includes a prediction algorithm for real-time localisation and tracking. To introduce more realistic mobility modelling, in \cite{RWP_Paper}, the traditional random waypoint model was extended by incorporating orientation of the user's equipment during both movement and pause times, resulting in improved performance in terms of reducing handover frequency, thus enhancing connectivity stability. Furthermore, an adaptive mobility prediction algorithm was developed in \cite{mobility+classes} for improving resource management efficiency while minimising operational costs.  In \cite{LB}, a fuzzy logic-based load-balancing scheme was introduced to address user mobility and light blockages, enhancing throughput with low computational complexity.}

Despite the works mentioned above and the extensive literature on OWC, optimising energy efficiency in laser and/or LED-based indoor OWC environments remains a significant challenge due to the practical and dynamic nature of indoor settings particularly in environments characterized by the unpredictable arrival of users and diverse demand profiles, which requires a flexible and adaptive optimisation approach. While the aforementioned studies have made significant strides in mobility management, resource allocation, and power optimisation, key challenges persist in integrating these components into a unified framework. 

Building on the foundations mentioned above, our work proposes a probabilistic demand prediction-optimised power allocation (PDP-OPA) framework that proactively predicts traffic demand and user arrival and departure to determine user association and dynamically allocate power at low complexity.  Our objective is to maximize the Consumption Factor (CF), which is defined as the maximum ratio of data rate to power consumption \cite{6522957}, while considering system and QoS constraints. To achieve this, we formulate a time discretised optimisation problem that mimics the dynamics of real-time and real-world environments. By proactively solving the optimisation problem for each time slot, our framework takes real-time decisions. For instance, during the period of time \( T_j \), the PDP proactively predicts the traffic demand and user distribution for the duration  \( T_{j+1} \) and determines the optical AP-user association. We then feed these information into a Lagrangian-based method developed to solve power allocation, incorporating dual variables to ensure constraints are satisfied while maximising the CF. 

This paper makes the following contributions:

\begin{itemize}

\item We model a  VCSEL-based OWC system that employs QAM-OFDM to enable high-speed data transmission designed to support mobile users with diverse demand requirements. Moreover, we integrate ZF precoding to mitigate interference, ensuring robust communication. Unlike conventional approaches, our system categorises users into $K$ demand classes, such as video streaming, voice communication and web browsing including Facebook and Instagram. This enables more adaptive resource allocation for varying service needs. 

\item For energy efficient OWC networks, a power allocation optimisation problem is formulated to maximise CF by jointly optimising user association and power allocation under the system constraints. To mimic real-time and real-world environments, we discretise time and this adds on to the complexity of the optimisation problem resulting from the coupling of the user association and power allocation. 

\item A key novelty of our approach is the development of the PDP-OPA framework specifically tailored for real-time and dynamic  OWC systems. Our proposed framework proactively predicts traffic demand and user arrival and departure to determine user association and dynamically allocate power whilst significantly reducing computational complexity and improving performance. For instance, during time slot $T_j$, the PDP predicts user demands and pre-determines AP-user associations for each demand class during the next duration $T_{j+1}$, allowing for proactive power allocation in dynamic environments.

\end{itemize}

The rest of the paper is organised as follows: Section \ref{system model} gives the system model, Section \ref{OP} details the formulation of the time discrete optimisation problem. Section \ref{Predictive and Proactive Strategy} details how user distributions and traffic demands are predicted proactively. Section \ref{Numerical Results and discussion} gives the results and finally Section \ref{Conclusion} gives the conclusions and recommendations for future work.

\section{System Model}
\label{system model}

\subsection{Room and system setup}
\begin{figure}[t]
\centering 
\includegraphics[width=\columnwidth]{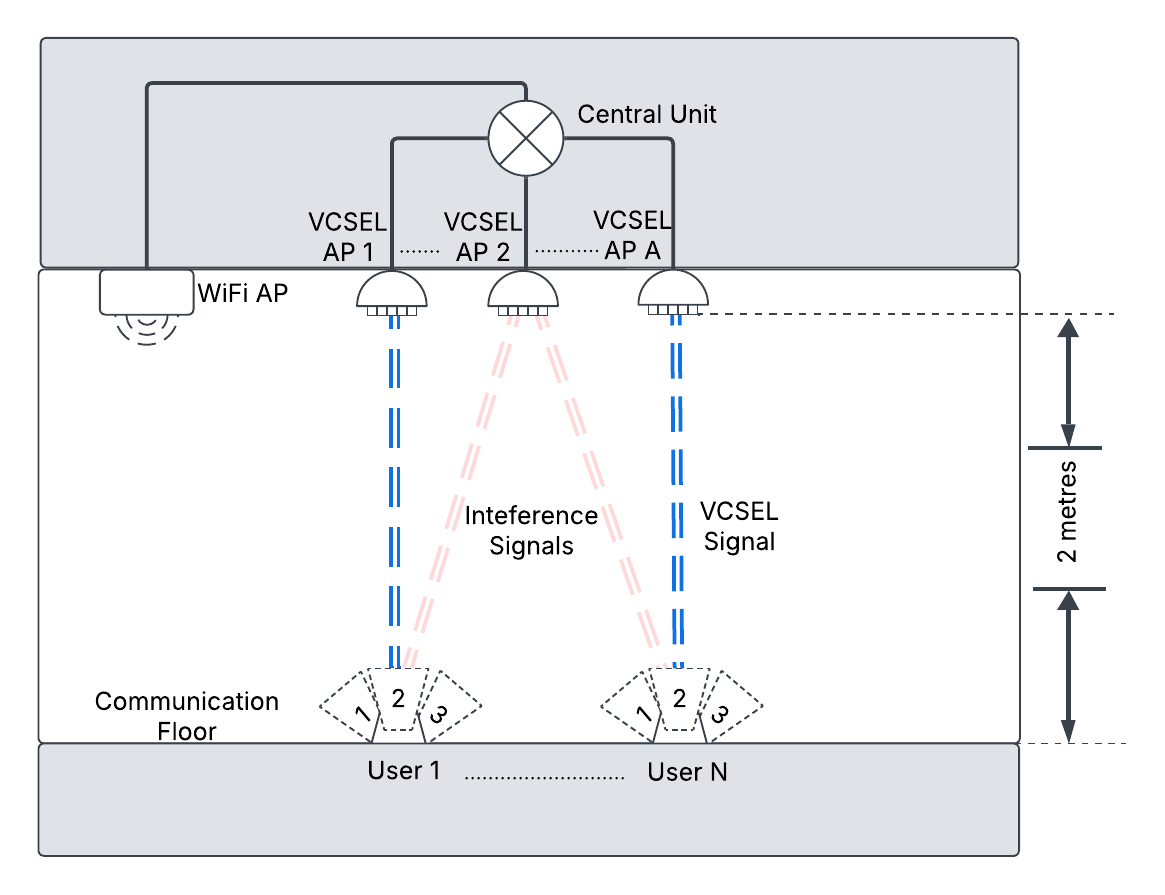} 
\caption{Laser-based OWC system model}
\label{fig:downlinksystemmodel}
\end{figure}

We consider an OWC system that uses VCSELs for data transmission in an indoor environment. Optical APs, denoted as $\mathcal{A}$, \( a = \{1 \ldots, A\} \), are strategically placed on the ceiling to ensure full room coverage, minimising interference and improving overall system performance. Each AP is composed of a $L_c \times L_c$ micro-lens array of VCSELs, transmitting data in a confined coverage area where mobile users arrive and depart as time progresses. In this system, all optical APs are interconnected with a WiFi AP deployed for uplink transmission, and to a central unit (CU) as shown in Fig. \ref{fig:downlinksystemmodel}, which plays a crucial role in coordinating APs and managing network resources.

The system is designed to support a number of users with time-varying demands denoted as $\mathcal{N}$, \( n = \{1, \ldots, N\} \), who subscribe to \( \mathcal{K} \), \( k = \{1,\ldots, K\} \), demand/service classes such as 8K Video Streaming,  Voice Communication, and General Browsing. Each user $n$ is equipped with an Angle Diversity Receiver (ADR), which is  a multi-element optical device, with \( N_{PD}=5 \) photodiodes, that uses Compound Parabolic Concentrators (CPCs) for maximum concentration gain. CPCs have  attractive characteristics such as the ability to efficiently collect light from a wide field of view, enhance signal power, and improve system resilience to misalignment and channel fading \cite{10299714,CPC_paper}.
Users of each service class denoted as $\mathcal{N}_k$, \( n = \{1, \ldots, N_k\} \),  are allocated power that is proportional to their respective class requirements, ensuring that high-demand applications receive sufficient power for a smooth user experience. Therefore,  each user $n$ subscribed to class $k$ is allocated power within the range \([{P_{\text{min}}}, {P_{\text{max}}}]\), where  ${P_{\text{min}}}$ guarantees the minimum Quality of Services (QoS) for that class. Conversely, ${P_{\text{max}}}$ represents the maximum power that can be allocated, considering system constraints, technical and optical AP limitations. In addition to this, each class has QoS requirements in the range $\geq {C_{\text{min},k}}$, where ${C_{\text{min},k}}$ is the minimum required QoS by subscribers of class $k$.

In order to preserve the dynamic nature of our system, characterised by mobile users, time-varying user demands and diverse demand/service classes, we take a structured approach as follows: We divide the serving time into fixed intervals, ${ T = \{T_1, \dots, T_j, T_{j+1}, \ldots, T_J\}}$, each of constant duration \( \tau \). As time progresses  from one time slot to another, the number and distribution of users changes, as well as their subscriptions to the service classes. 
\subsection{Downlink System Modelling}

This section outlines the downlink communication system, detailing VCSEL-based APs, optical channel gain, signal transmission using DCO-OFDM with adaptive QAM, ZF for interference mitigation, and the calculation of user data rates.

\subsubsection{VCSEL Access Points}
We use VCSEL emitters for data transmission owing to the numerous advantages that they offer such as high bandwidth, energy and power efficiency, compact size that allows easy integration into small transmitters, and their circular beam shape and narrow divergence angles that improve optical coupling and link quality \cite{qidan_towards_2022,Qidan24,10647466,9844102,10325483,9803253,eye_safefy_book}. We divide the coverage area into distinct zones, each serviced by a dedicated optical AP comprised of a ${L_{c} \times L_{c}}$ VCSEL micro-lens array optical AP. Each AP sends different signals to individual users. We assume that the VCSEL exhibits a Gaussian intensity profile with two key parameters: the beam waist \( w_0 \), and the wavelength \( \lambda \). Assuming the beam propagates along the $z$-axis, the intensity distribution at a point on the transverse plane at the receiver is given by \cite{9217158}:
\begin{equation}
I(r,z) = \frac{2 P_{\text{out}}}{\pi w^2(z)} \exp \left(-\frac{2 r^{2}}{w^2(z)} \right),
\label{eq1}
\end{equation}

\noindent where \( P_{\text{out}} \) is the optical power emitted by a single VCSEL, and \( r \) and \( z \) represent the radial and axial positions, respectively. The beam radius \( w(z) \) evolves along the propagation axis as:

\begin{equation}
w(z) = w_0 \sqrt{1 + \left( \frac{z}{z_R} \right)^2},
\label{eq2}
\end{equation}

\noindent where \( w_0 \) is the beam waist at the origin, and \( z_R \) is the Rayleigh range defined as $z_R = \frac{\pi w_0^2 n}{\lambda}$, where \( n \) is the refractive index of the medium. The addition of a micro lens modifies the VCSEL output beam but still maintains the Gaussian profile. We use the ABCD transformation \cite{ABCD}, a well-used method for describing the effect of an optical element on a laser beam passing through it \cite{9803253}, to model the micro-lens' effect on the VCSEL output beam, and as a result of the transformation, the beam waist becomes minimum at distance $d_2$ after the lens. This is calculated as follows \cite{10621108}:

\begin{equation}
d_2 = \frac{\left(\frac{1}{f} - \left(1 - \frac{d_1}{f}\right) d_1 \left(\frac{\lambda}{\pi w_0^2}\right)^2\right)}{\left(\frac{1}{f^2} + \left(1 - \frac{d_1}{f}\right)^2 \left(\frac{\lambda}{\pi w_0^2}\right)^2\right)},
\label{eq3}
\end{equation}
\noindent where \(d_1\) is the distance between the VCSEL and the lens, $f$ is the focal length and $\lambda$ is the VCSEL wavelength. Moreover, the new minimum waist $w_d$ at a distance $d_2$ after the lens is calculated as \cite{10621108}:

\begin{equation}
w_d = \frac{\lambda f}{\pi w_0 \sqrt{1 + \left(1 - \frac{d_2}{f}\right)^2 \frac{\lambda^2}{\pi^2 w_0^4} f^2}}.
\label{eq4}
\end{equation}
The beam divergence, $\theta$, after the lens is given by $\theta_2 = \frac{\theta}{k}$, where $k$ is the magnification factor of the lens, i.e., $k=  \frac{w_d}{w_0}$. The transmitted optical power of the transformed beam from a single VCSEL, $P_{v}$, within a circular radius $r_0$ at the receiver plane is given by:

\begin{equation}
P_{v} = P_{\text{out}} \left(1 - \exp\left(-\frac{2 r_0^2}{(w_d(z))^2}\right)\right),
\label{eq5}
\end{equation}
where $w_d(z)$ is the beam radius at distance $z$, representing the propagation distance from the source to the receiver. The exponential term $\exp\left(-\frac{2 r_0^2}{(w_d(z))^2}\right)$ accounts for the Gaussian spatial distribution of the optical power over the receiver plane, with $1 - \exp\left(-\frac{2 r_0^2}{(w_d(z))^2}\right)$ representing the fraction of the total power enclosed within the radius $r_0$. The total transmitted power from a VCSEL array optical AP, $a$, is the sum of the output power from all the individual VCSELs in the array, expressed as:

\begin{equation}
P_{a} = \sum_{i=1}^{L_c \times L_c} P_{i,v},
\label{eq6}
\end{equation}

\noindent where $P_{i,v}$  the transmitted power from the $i$-th VCSEL in the array. It should be noted that eye safety considerations are based on the total output power from the optical AP, $P_a$, requiring that power levels for each VCSEL are carefully managed to ensure the total emitted power remains within safe limits for human exposure. As a result $P_{i,v} \leq P_{i,\text{safe}}$ where $P_{i,\text{safe}}$ is the maximum power per VCSEL for eye safety and is calculated as \cite{9803253}:
 
\begin{equation}
P_{i,\text{safe}} = \frac{1}{\eta} \mathrm{MPE}(\pi r_p^2),
\label{eq007}
\end{equation}

\noindent where \cite{9803253}:

\begin{equation}
\eta = 1 - \exp\left(-\frac{2 r_p^2 (\pi w_o^2)^2}{(\lambda \times \mathrm{MHP})^2}\right),
\label{eq28}
\end{equation}

\noindent and $\mathrm{MHP}$ is the most hazardous position which is defined as the location at which the ratio of the exposure level to the $\mathrm{MPE}$ value is at its maximum. $r_p$ is the radius of the eye pupil. We consider that the pupil can be as large as 6–7 mm when it is wide open \cite{9217158}. 

\subsubsection{Optical Channel Gain}

The DC gain between the \( i \)-th VCSEL and the \( PD \)-th photodiode of user $n$ is given by \cite{LoS_Haas}: 

\begin{align}
H_{iPD}^n &= \frac{2 N_{PD} A_{PD} G}{\pi w_d^2(d_{na} \cos \phi_{ni})} 
\exp \left( -\frac{2 d_{na}^2 \sin^2 \phi_{ni}}{w_d^2(d_{na} \cos \phi_{ni})} \right) \nonumber \\
&\quad  \cos \psi_{iPD} \chi_{\theta_{CPC}} (\psi_{iPD}),
\end{align}

\noindent where  \( A_{PD} \) is the percentage active area of the photodetector. \( G \) is the gain of the concentrator, \( d_{na} \) is the Euclidean distance between the \( n \)-th user and the transmitter, \( \phi_{ni} \) is the radiance angle between the \( i \)-th VCSEL beam and the \( n \)-th user's position. \(\psi_{iPD} \) is  the incidence angle of the beam relative to the normal of the receiver plane $PD$-th receiver element. $\chi_{\theta_{CPC}} (\psi_{iPD})$ is an indicator function that ensures the beam contributes to the gain only if the incidence angle \( \psi_{iPD} \) satisfies $0 \leq \psi_{iPD} \leq \theta_{acc}$, otherwise, the gain is zero.

\subsubsection{Signal Transmission}

In this system, we consider DCO-OFDM combined with adaptive QAM in order to achieve high spectral efficiency as in \cite{10299714}. The system leverages direct current biasing to ensure non-negative signal values required for driving the VCSELs. The DCO-OFDM system is based on an $M$-point IFFT at the transmitter and $M$-point FFT at the receiver. At the transmitter, the baseband bandwidth of the system is equally divided into $M$ orthogonal sub-channels, and binary data is mapped according to an F-QAM constellation onto a total of $\frac{M}{2} - 1$ data carrying sub-carriers. Each sub-carrier is assigned bits, which are mapped to a complex QAM sub-symbol such that the bits of the $l^{th}$ sub-carrier are mapped onto the $l^{th}$ complex QAM sub-symbol. That is, 
 \(X_{M-l} = {X}_l^*\) for \(l = 1,  \dots, \frac{M}{2} - 1\), and \(X_{0} = X_{\frac{M}{2}}=0\) and $[]^*$ is conjugate operator \cite{10299714}.

 In the frequency domain, applying an $M$ point IFFT to the DCO-OFDM frame can transform  the $M$ elements of  $\mathbf{X}= [X]_{l=0, 1,\dots, M-1}$ into $M$ real samples in the time domain. Then, the output signal is \cite{10299714}:
\begin{equation}
x(t) = \frac{1}{\sqrt{M}} \sum_{l=0}^{M-1} \alpha {X}_l e^{q \frac{2\pi lt}{M}}, 
\label{eq7}
\end{equation}
where, $t = 0, 1, \dots, M-1$ and $q = \sqrt{-1}$. Prior to transmission, the time-domain signal is power-normalized. The normalization factor \( \alpha \), defined as $\alpha =\sqrt{ \frac{M}{M-2}}$, ensures that the average power of the time-domain signal is normalized to unity i.e $E\left[ |x(t)|^2 \right] = 1$.

A DC bias, \(I_{\text{DC}}\), must be added to the signal to ensure all values remain non-negative, meeting the requirements of the optical source. The DC bias  is chosen to position the signal's peak-to-peak amplitude within the linear range of the optical device, preventing nonlinear distortion and minimizing clipping.  The transmitted signal is then  given by 

\begin{equation}
x_{\text{op}}(t) = \sqrt{P} x(t) + I_{\text{DC}}.
\label{eq8}
\end{equation}
where ${P}$ is the power allocated to transmit  $x(t)$.

In our VCSEL-based system, the BER for a QAM scheme with an $F$ size constellation is tightly upper-bounded as in \cite{9217158} by the following expression, which is accurate to within 1~dB for \(F \geq 4\) and $\mathrm{SINR}$ values \(0 \leq \mathrm{SINR} \leq 30\)~dB \cite{BER}:

\begin{equation}
\mathrm{BER} \leq 0.2\, \exp\!\left(-\frac{1.5\,\mathrm{SINR}}{F-1}\right).
\label{BER_eq1}
\end{equation}
In (\ref{BER_eq1}), the BER decreases exponentially with increasing SINR, and is also dependent on the QAM constellation size \(F\). A larger constellation (i.e., more symbols) results in a higher BER at a given SINR because the symbols are more closely spaced. To ensure that all direct channels meet the same BER requirement in an adaptive QAM system, the modulation order is dynamically adjusted. By setting the upper bound in the BER expression equal to the target BER and solving (\ref{BER_eq1}) for \(F\), we determine the maximum constellation size that can be supported on a given channel. This results in \cite{9217158}:

\begin{equation}
F= 1 + \frac{\mathrm{SINR}}{\Gamma}, 
\label{BER_M_eq}
\end{equation}

\noindent where $\Gamma$ is the SINR gap due to the required bit error rate (BER) performance before forward error correction (FEC) and is calculated as in \cite{10299714,BER}:

\begin{equation}
\Gamma = \frac{-\ln{5\text{BER}}}{1.5},
\label{gamma_eq}
\end{equation}
Once $F$ is determined, the number of bits transmitted per channel use for each sub-carrier is equal to $\mathrm{\log_2 M}$.

\subsubsection{Zero Forcing}

We consider ZF to mitigate multi-user interference (MUI) effectively within the coverage area of each optical AP. Note that, the total number of users in the network is \(N\), and each optical AP serves $\mathcal{N}_a$, \( n = \{1, \ldots, N_a\} \), users subscribing to different demand classes. Considering \(M\) orthogonal channels, with the signal allocated to \( L \), \( l = \{1,\ldots, \frac{M}{2} - 1\} \) data carrying subcarriers,  the received signal per subcarrier \(l\) in the time domain  can be expressed as:

\begin{equation}
\textbf{y}(t) =  \textbf{H}    \textbf{g} (t) +\textbf{ I} (t) + \boldsymbol{v}(t),
\label{eq9}
\end{equation}

\noindent where:

\begin{itemize}
    \item \(\textbf{y}(t)\) is the received FD symbol vector,
    \item \(\textbf{H} \in (\mathbb{R})^{N_a}\) is the channel gain matrix between the \(N_a\) users and the AP,
    \item \(\textbf{ I} (t)\) is the interference imposed on the users,
    \item \(\boldsymbol{v}(t)\) is the receiver noise vector. 
\end{itemize}

 The transmitted signal on the \(l\)-th subchannel is formulated as; $\textbf{g}(t) = \textbf{W} \cdot \textbf{X} (t)$, where \(\textbf{X}(t) \in (\mathbb{C})^{N_a}\) is the multi-user FD symbol vector, and the ZF precoding matrix \(\textbf{W} \in (\mathbb{C})^{N_a}\) is explicitly given by the pseudo-inverse of the channel matrix \(\textbf{H}\), i.e., $\textbf{W} = \textbf{H}^+$, where \(\textbf{H}^+\) is the pseudo-inverse of \(\textbf{H}\). Therefore, the received information can be decomposed into \(|N_a|\) parallel streams. From (\ref{eq9}), 
 the received signal per subcarrier $l$  in the time domain for a specific user \(n \in \mathcal{N}_a\)  subscribing to class $k$ is given by
 
\begin{equation}
y^{n}_{a,k} (t)= R_{\text{PD}} H_{a}^n  \sqrt{P^{n}_{a,k}}   x^{n}(t)+ I^{n} (t) + v^{n} (t),
\label{eq10}
\end{equation}

\noindent where:

\begin{itemize}
\item $R_\text{PD}$ is the responsivity,
\item $H_{a}^n= \sum^{5}_{PD=1} \sum^{L_{c} \times L_{c}}_{i=1}H_{iPD}^n$
\item $P_{\text{a, k}}^{n}$ is the power allocated to user $n \in \mathcal{N}_a$ to transmit $x^{n}(t)$, which must be determined to ensure high QoS, i.e., a user rate is $\geq {C_{\text{min},k}}$,
\item $I^{n} (t)$ is the interference imposed on user $n$,
\item \(v^{n} (t)\) is the noise on user $n$.
\end{itemize}

Following detection, the DC bias is removed, and the signal is passed through an M-point FFT to recover the data sub-carriers in the frequency domain. The FFT of the received signal in (\ref{eq10}) is expressed as:

\begin{equation}
{Y}^{n}_{a,k}= \frac{1}{\sqrt{M}} \sum_{t=0}^{M-1} y^{n}_{a,k} (t) e^{-q \frac{2\pi lt}{M}},
\label{eq11}
\end{equation}

\subsubsection{Achievable Rate}

As in \cite{10299714}, we consider a low-pass system with a flat frequency response, primarily constrained by the receiver’s bandwidth, given that VCSELs have a wide modulation bandwidth exceeding 30 GHz. As a result, all OFDM subcarriers experience uniform channel gain, denoted as \(\textbf{H}_l = \textbf{H}_0\) for \(l = 1, \dots, M-1\). 

When employing adaptive QAM modulation, the achievable rate for the $l$th data carrying subcarrier can be expressed as:

\begin{equation}
\begin{split}
      C_l = \frac{1}{T} \log_2 \left( 1 + \frac{\mathrm{SNR}_l}{\Gamma} \right), 
\end{split}
\label{eq12}
\end{equation}

The received $\mathrm{SNR}$ per sub-carrier can be easily derived from (\ref{eq10}) and expressed as \cite{10299714, Mingqing_EE}:

\begin{equation}
\begin{split}
\mathrm{SNR}_l = \frac{R_{\text{PD}}^{2} \alpha^{2} P}{ \xi {\sigma}^2_n}.
\end{split}
\label{eq13}
\end{equation}
It should be noted that \(\xi = \frac{M-2}{M}\) represents the sub-carrier utilisation ratio, reflecting that noise is only observed on the $\frac{M}{2}-1$ data-carrying sub-carriers and $\alpha^2 = \frac{M}{M-2} = \xi^{-1}$. Therefore from (\ref{eq12}) and (\ref{eq13}),  the total achievable data rate for the DCO-OFDM system can be expressed as:

\begin{equation}
\begin{split}
     C_{total} = \sum_{l=1}^{\frac{M}{2} - 1} \frac{2B}{M} \log_2 \left( 1 + \frac{\mathrm{SINR}_l}{\Gamma} \right),
\end{split}
\label{eq14}
\end{equation}
where B is the bandwidth. The baseband OFDM signal spans the available spectrum from \(-B\) to \(B\), giving a sub-channel bandwidth of $\frac{1}{T} = \frac{2B}{M}$.
From the multi-user scenario in equation (\ref{eq9}), and from equations (\ref{eq10}-\ref{eq14}),  the user data rate, $C_{a,k}^n$, for user $n$  subscribing to class $k$ and served by AP $a$, can be expressed as:

\begin{equation}
\begin{split}
C^{n}_{a,k}
= \xi B \,\log_{2}\Biggl( 
1 \;+\; \frac{\exp(1)}{2\pi} \cdot \,\frac{ (R_{\text{PD}}H_{a}^n)^{2}\,P_{a,k}^{n}}
               {\Gamma\!\Bigl(\,\xi^{2}\,\sigma_{n}^{2} 
                  \;+\; \sum_{\substack{a' \in A \\ a' \neq a}} P_{a',k}^{n}\Bigr)}
\Biggr).
\end{split}
\label{eq15}
\end{equation}
It is worth mentioning that the user  date  rate is is determined using the lower bound of the Shannon capacity formula as in \cite{Data_Rate_Paper}, accounting for the characteristic of the optical signal, i.e., IM/DD. In \eqref{eq15},
the noise is a white Gaussian noise with zero mean and a variance of $\sigma^2_n = N_TB$, where $N_T$ is noise power spectral density, and is calculated as in our previous work \cite{10621108}. 

\section{ Energy Efficiency Optimisation Problem}
\label{OP}
{Based on the multi-user OWC system model presented in
Section \ref{system model}, we now formulate a discrete-time EE optimisation problem under power allocation, user association, and QoS constraints. Our  aim is to maximise the system's CF for each time period due to the fact that the network is dynamic and  changes in the network conditions are frequent as time progresses. Therefore, the EE optimisation problem for the time period $T_{j+1}$ is given as:} 

\begin{equation}
\max_{S^{T_{j+1}}_{a,k,n}, P_{{a, k}}^{n}} \frac{{\sum_{k \in \mathcal{K}} \sum_{n \in \mathcal{N}_k} \sum_{a \in \mathcal{A}}} S^{T_{j+1}}_{a,k,n} \log  \big(C^{n}_{a,k} (T_{j+1})\big)}{{\sum_{k \in \mathcal{K}} \sum_{n \in \mathcal{N}_k} \sum_{a \in \mathcal{A}}} S^{T_{j+1}}_{a,k,n} P_{{a, k}}^{n} },
\label{eq24}
\end{equation}
where $C^{n}_{a,k} (T_{j+1})$ is the lower bound of the capacity for user \(n\) subscribing to class $k$ and potential served by AP $a$ during the time period $T_{j+1}$. Moreover, $S^{T_{j+1}}_{a,k,n}$ is the association variable for user \(n\) to determine whether or not user \(n\) is allowed service $k$ by AP \(a\) during the time period $T_{j+1}$. The constraints of the optimisation problem are listed as follows:
\begin{equation}
\sum_{a \in \mathcal{A}} S^{T_{j+1}}_{a,k,n} \leq 1, \quad \forall k \in \mathcal{K}, \forall n \in \mathcal{N}_{k},
\label{eq25}
\end{equation}

\begin{equation}
C^{n}_{a,k} \geq C_{\text{min},k}^n, \quad \forall k \in \mathcal{K}, \forall n \in \mathcal{N}_{k}, \forall a \in \mathcal{A},
\label{eq26}
\end{equation}

\begin{equation}
\sum_{k \in \mathcal{K}} \sum_{n \in \mathcal{N}_k} S^{T_{j+1}}_{a,k,n} P_{{a, k}}^{n} \leq P_{{a}}, \quad \forall a \in \mathcal{A},
\label{eq27}
\end{equation}

\begin{equation}
P_{\text{min}} \leq P_{{a, k}}^{n} \leq P_{\text{max}}, \quad  \forall k \in \mathcal{K}, \forall n \in \mathcal{N}_{k}, \forall a \in \mathcal{A},
\label{eq28}
\end{equation}

\begin{equation}
P_{{a, k}}^{n} \geq 0,  S^{T_{j+1}}_{a,k,n} \in \{0,1\}, \forall k \in \mathcal{K}, \forall n \in \mathcal{N}_{k}, \forall a \in \mathcal{A}.
\label{eq30}
\end{equation}
Firstly, the constraint in (\ref{eq25}) ensures that each user $n$ subscribing to class $k$ is served by no more than one AP during  each time period, and the constraint in (\ref{eq26}) ensures high QoS, where \(C_{\text{min},k}^n\) is the minimum data rate required to meet the QoS for user $n$ subscribing to class $k$. Moreover, the constraint in (\ref{eq27}) ensures that the total transmit power allocated by each optical AP $a$ to all users it serves, in all user classes it provides, does not exceed the optical AP's maximum allowable power $P_a$. It is worth mentioning that $P_a$ is limited due to the eye safety constraints and determined using equation (\ref{eq007}), and  constraint (\ref{eq27}) also helps measuring the level of interference imposed on the users of each optical AP $a$ from the adjacent APs, i.e., $P_{{a', k}}^{n}, a' \neq a $, which is limited due to the use of laser beams, VCSELs, for data transmission\cite{Qidan24,10325483}. Furthermore, the constraint in (\ref{eq28}) ensures that the received power \(P_{\text{a, k}}^{n}\) allocated to each user \(n\) subscribing to class \(k\) and served optical AP \(a\) remains within a defined range, bounded by \(P_{\text{min}}\) and \(P_{\text{max}}\). 

Lastly, the constraint in (\ref{eq30}) denotes the feasible region of the formulated optimisation problem.

The optimisation  problem in (\ref{eq24}) has a non-convex objective function and is NP-hard  due to the coupling of \(S_{a,k,n}^{T_{j+1}}\) and power allocation, and it has approximate complexity of $\mathcal{O}(A^{N})$. Moreover, the modelled network is highly dynamic where changes in user's distribution, number, and  traffic demand occur as  time progresses. Therefore, solving the optimization problem for the time period $T_{j+1}$ of  \( \tau \) duration is infeasible. 
For valid and up-to-date solutions, during each time interval, ${T_j}$, our network must capture the changes in the network for the next time slot ${T_{j+1}}$. Then, two sub-problems for user-AP association and power allocation must be solved proactively to ensure continuous adaptation and  performance enhancement. Let \( \mathcal{S} \) denote the user association vector, and \( \mathcal{P} \) denote the power allocation vector. These vectors must  be optimised alternately, until a stationary point is obtained. Upon convergence of the iterative algorithm, the solutions to both sub-problems fulfil the Karush-Kuhn-Tucker (KKT) \cite{KKT_book} conditions for the initial optimisation problem.

\section{Predictive and Proactive Strategy }
\label{Predictive and Proactive Strategy} 
In this section, a PDP algorithm  is proposed for real-time adaptation to the changes in the network. Recall that, the time  is divided into periods ${ T = \{T_1, \dots, T_j, T_{j+1}, \ldots, T_J\}}$, and therefore,  the PDP must use data from each time period,  \( T_j \), to predict the number of users, and their distribution among the optical APs and the demand classes for the next period \( T_{j+1} \). After that, user association and power allocation are determined by our full proposed PDP-OPA strategy. We consider that  the  arrival process of new and handoff users follows a Poisson process  as in \cite{main_PDP_paper}, and  the prediction process considers user arrivals and departures, average class holding time, and the probability distribution of the class holding time.  We focus on a certain optical AP, $a \in \mathcal{A}$, and on a single demand class, $k \in \mathcal{K} $, for simplicity of presentation. Note that, similar calculations apply to the other optical APs and demand classes. We carry out the proposed PDP-OPA in the following steps:
 \subsection{User Distribution}
\label{user_mobility_section}
Each AP tracks the number of users  subscribing to demand class \(k\) within its coverage area during each time period, forming a time vector \( \tilde{T}_j^k \). The number of users  \( N_a^k(t^{o}_{j}) \), where \( o = \{1, 2, \ldots, \lvert \tilde{T}_j^k \rvert \} \),  present at any given time, $t^{o}_{j}$, is used to predict the number of users at a future time \( t^{o}_{j} + \tau \), where \( \tau \) is the prediction duration. We employ a prediction method similar to that in \cite{PDP_paper_2} that follows a transient distribution model of the \( M/G/\infty \) system, which considers both user arrivals and residence times within an AP’s coverage area and in each class. The probabilities \( p_\tau^{(a,k)} \) and \( q_\tau^{(a,k)} \) represent the likelihood of users remaining and new users arriving in the coverage area of AP $a$ with a subscription to class $k$, respectively, during the interval \([t^{o}_{j}, t^{o}_{j} + \tau]\).

In an indoor environment with services such as on-demand video streaming, session durations follow a heavy-tailed distribution \cite{main_PDP_paper,streaming}. This is a probability distribution in which extreme values (i.e., very long durations) occur with a non-negligible probability, meaning that the distribution has a high variance and a slow decay in its tail. Thus, while most users remain for a short period, there is a significant probability of extended session durations. Since heavy-tailed distributions are mathematically complex, we approximate them using a two-stage hyper-exponential distribution, which consists of two exponential components with different rates \cite{PDP_paper_2, main_PDP_paper}. This allows for an approximation of the short long duration behaviour of the session durations separately.

We further characterise user distribution by the user residence time, $T_{n,r}^{(a,k)}$, which follows an exponential distribution with mean $\tilde{T}_{n,r}^{(a,k)}$. Moreover, the holding time is also defined as $T_{n,h}^{(a,k)} = \min(T_{n,d}^{(a,k)}, T_{n,r}^{(a,k)})$, where \( T_{n,d}^{(a,k)} \) is the session duration time for a a certain user $ n$ served by AP $a$ with a class $k$ traffic demand. As in \cite{PDP_paper_2} and  \cite{main_PDP_paper}, we determine the Probability Distribution Function (PDF) for the session duration time with mean \( \tilde{T}_{n,d}^{(a,k)} \) by:

\begin{align}
f_{T_{n,d}^{(a,k)}}(t) &= \frac{\omega}{\omega + 1} \cdot \frac{\omega}{\tilde{T}_{n,d}^{(a,k)}} \cdot e^{-\frac{\omega}{\tilde{T}_{n,d}^{(a,k)}} t} \notag \\
&\quad + \frac{1}{\omega + 1} \cdot \frac{1}{\omega \tilde{T}_{n,d}^{(a,k)}} \cdot e^{-\frac{1}{\omega \tilde{T}_{n,d}^{(a,k)}} t}, \omega \geq 1, \quad t \geq 0 
\label{eq19}
\end{align}$\omega$ is a shape parameter used to model session durations and holding times. It determines the weighting of the two exponential components in the distribution, capturing the variability and heavy-tailed behaviour of real-world session durations. Larger $\omega$ values increase the likelihood of shorter durations, while smaller $\omega$ values allow for longer durations, reflecting heavy-tailed characteristics like the one in (\ref{eq19}). The PDF of \( T_{n,h}^{(a,k)} \) is given as \cite{PDP_paper_2, main_PDP_paper}:

\begin{align}
f_{T_{n,h}^{(a,k)}}(t) &= \left(\frac{\omega}{\omega + 1} \right) \cdot \left(\frac{1}{\tilde{T}_{n,r}^{(a,k)}} + \frac{\omega}{\tilde{T}_{n,d}^{(a,k)}}\right) \cdot \notag \\
&\quad e^{-\left(\frac{1}{\tilde{T}_{n,r}^{(a,k)}} + \frac{\omega}{\tilde{T}_{n,d}^{(a,k)}}\right) t} + \left(\frac{1}{\omega + 1}\right) \cdot \notag \\ 
&\quad \left(\frac{1}{\tilde{T}_{n,r}^{(a,k)}} + \frac{1}{\omega \tilde{T}_{n,d}^{(a,k)}}\right) \cdot e^{-\left(\frac{1}{\tilde{T}_{n,r}^{(a,k)}} + \frac{1}{\omega \tilde{T}_{n,d}^{(a,k)}}\right) t}, \notag \\
&\quad \omega \geq 1, \quad t \geq 0 
\label{eq20}
\end{align}
\subsection{ Predicting User Demand}
As mentioned above, at time \( t^{o}_{j} \in \tilde{T}_j^k \) in the time period $T_j$,  the number of active users \( N_a^k(t^{o}_{j}) \) is used to predict the user demand probabilistically at a future time \( t^{o}_{j} + \tau \) in the next time period \( T_{j+1} \). Here, \( \tau \) represents the prediction interval, and the predicted demand is denoted by \( \tilde{N}_a^k(t^{o}_{j} + \tau) \). The probability distribution of \( N_a^k(t^{o}_{j} + \tau) \), conditioned on \( N_a^k(t^{o}_{j}) \), forms the basis for estimating \( \tilde{N}_a^k(t^{o}_{j} + \tau) \), {where \( N_a^k(t^{o}_{j} + \tau) \) is the number of active users at time $(t^{o}_{j} + \tau)$}. This estimate can be expressed using a design parameter, \( \epsilon_a^k \), defined as follows \cite{PDP_paper_2}:
\begin{align}
   & P_r\Big(N_a^k(t^{o}_{j} + \tau) > \tilde{N}_a^k(t^{o}_{j} + \tau) \mid N_a^k(t^{o}_{j})\Big) \leq \epsilon_a^k, \nonumber\\
    &\quad \forall a \in \mathcal{A}, \forall k \in \mathcal{K}
      \label{eq33}
\end{align}

\noindent where \( \epsilon_a^k \in [0, 1] \) denotes the probability that \( N_a^k(t^{o}_{j} + \tau) \) exceeds \( \tilde{N}_a^k(t^{o}_{j} + \tau) \). The calculation of \( \tilde{N}_a^k(t^{o}_{j} + \tau) \) relies on the conditional Probability Mass Function (PMF) of \( N_a^k(t^{o}_{j} + \tau) \) given \( N_a^k(t^{o}_{j}) \), denoted \( P_{N_a^k(t^{o}_{j} + \tau) \mid N_a^k(t^{o}_{j})}(o) \). Assuming user arrivals follow a Poisson process, with class holding times having a general distribution, and users served without queuing, the transient distribution of the M/G/\(\infty\) model is used to compute \( P_{N_a^k(t^{o}_{j} + \tau) \mid N_a^k(t^{o}_{j})}(o) \) \cite{MANDJES2011507}.

We define the following parameters for a stationary user arrival and departure process:
\begin{itemize}
    \item \( p_{\tau}^{(a,k)} \) - The probability that a user in class \( k \) present at \( t^{o}_{j} \) remains in the same user class at \( t^{o}_{j} + \tau \). This represents the likelihood that a user who is currently active in class $k$ remains active at the next time step in the same class. It depends on the user's session duration and mobility. Users engaged in longer video streaming sessions are more likely to persist than users browsing briefly.
    \item \( q_{\tau}^{(a,k)} \) - The probability that a user arriving in class \( k \) during \( (t^{o}_{j}, t^{o}_{j} + \tau) \) remains at \( t^{o}_{j} + \tau \). This reflects the probability that a new user arriving during the current time step will still be active at the next time step. It depends on the average class holding time and user behavior. A high $ q_{\tau}^{(a,k)}$ implies users tend to stay connected soon after arriving, common in stationary users.
    \item \( Y_B(\iota, \beta) \) - A binomial random variable with parameters \( \iota \) and \( \beta \),
    \item \( Y_P(\beta) \) - A Poisson random variable with mean \( \beta \).
\end{itemize}

The demand at \( t^{o}_{j} + \tau \), given \( N_a^k(t^{o}_{j}) \), is modelled as:
\begin{equation}
    N_a^k(t^{o}_{j} + \tau) =_d Y_B \left(N_a^k(t^{o}_{j}), p_\tau^{(a,k)} \right) + Y_P \left(\mu_\tau^{(a,k)} \tau q_\tau^{(a,k)} \right),
    \label{eq34}
\end{equation}
where \( =_d \) denotes equality in distribution, and \( \mu_\tau^{(a,k)} \) is the arrival rate of new or handoff users in class \( k \). For an optical AP \( a \), \( \mu_\tau^{(a,k)} \) is determined by counting new arrivals to class \( k \) and dividing by the elapsed time.

The probabilities \( p_\tau^{(a,k)} \) and \( q_\tau^{(a,k)} \) are defined by \cite{PDP_paper_2}:

\begin{equation}
p_{\tau}^{(a,k)} = \frac{1}{E[T_{n,h}^{(a,k)}]} \int_{\tau}^{\infty} (1 - F_{T_{n,h}^{(a,k)}}(s)) \, ds,
\label{eq36}
\end{equation}

\begin{equation}
q_{\tau}^{(a,k)} = \frac{E[T_{n,h}^{(a,k)}]}{\tau} (1 - p_{\tau}^{(a,k)})
\label{eq39}
\end{equation}

\noindent where \( E[T_{n,h}^{(a,k)}] \), the average class holding time, is given by:

\begin{equation}
\begin{aligned}
E[T_{n,h}^{(a,k)}] &= \left(\frac{\omega}{\omega + 1}\right) \left(\frac{1}{\frac{1}{T_{n,r}^{(a,k)}} + \frac{\omega}{T_{n,d}^{(a,k)}}}\right) \\
&\quad + \frac{1}{\omega + 1} \cdot \frac{1}{\frac{1}{T_{n,r}^{(a,k)}} + \frac{1}{T_{n,d}^{(a,k)} \cdot \omega}},
\end{aligned}
\label{eq40}
\end{equation}

\noindent and \( F_{T_{n,h}^{(a,k)}}(s) = \int_0^s f_{T_{n,h}^{(a,k)}}(t) \, dt \) is the CDF of \( T_{n,h}^{(a,k)} \). The variable $s$ represents time during the prediction interval for which the probability of class holding time \( T_h^{(n,k)} > s \) is being evaluated. These probabilities are then weighted to determine the persistence (\( p_{\tau}^{(n,k)} \)) and arrival (\( q_{\tau}^{(n,k)} \)) probabilities. Using Equations (\ref{eq34})-(\ref{eq39}), \(P_{N_a^k(t^{o}_{j} + \tau) \mid N_a^k(t^{o}_{j})}(o) \) can be found, allowing \( \tilde{N}_a^k(t^{o}_{j} + \tau) \) to be determined using (\ref{eq33}) as the minimum integer satisfying:
\begin{equation}
    \sum_{i=0}^{\tilde{N}_a^k(t^{o}_{j} + \tau)} P_{N_a^k(t^{o}_{j} + \tau) \mid N_a^k(t^{o}_{j})}(o) \geq (1 - \epsilon_a^k), \quad \forall a \in \mathcal{A}, \forall k \in \mathcal{K}
\end{equation}
\subsection{Storing Predicted Values}
Once the prediction process is concluded, the predicted values  for all user classes, i.e., \( \tilde{N}_a^k(t^{o}_{j} + \tau) \), within the serving area of optical AP \( a \) are stored by the controller in a vector \( \hat{N}^{a,k}_{(j+1)} \), which represents the predicted demand during the next time period \( T_{j+1} \).

\subsection{User Association}
The controller, with knowledge of the predicted demand \( \tilde{N}_a^k(T_{j+1}) \) from \( \hat{N}^{a,k}_{(j+1)} \)  for every \( k \in \mathcal{K} \) within the coverage region of each optical AP $a$, solves the corresponding user association problem at each AP to determine the binary assignment variable $S^{T_{j+1}}_{a,k,n}$ for all users. Therefore, the user association vector \( \mathcal{S}_{(j+1)} \) for users during \( T_{j+1} \) can be determined. Based on the user association vector \( \mathcal{S}_{(j+1)} \), each AP \( a \) can determine the maximum number of users in class \( k \) that can be supported by this AP during \( T_{j+1} \) given \( \mathcal{P}_{(j+1)} \).

\subsection{Power Allocation}
Using the user association vector \( \mathcal{S}_{(j+1)} \),  the power allocation sub-problem can be solved  to determine the corresponding power allocation vector \( \mathcal{P}_{(j+1)} \) by applying full dual decomposition \cite{1664999}, leveraging Lagrangian multipliers. To do this first, the energy efficiency  per optical AP for the time period \( T_{j+1} \) is determined during time slot $T_j$ as follow:

\begin{align}
 {\text{EE per optical AP} = \max_{P_{a,k}^{n}} EE},   
\end{align}

\noindent where the energy efficiency is given by:

\begin{align}
EE = \frac{\sum_{k \in \mathcal{K}} \sum_{n \in \mathcal{N}_{a,k}} U(C^{n}_{a,k}(T_{j+1}))}{\sum_{k \in \mathcal{K}} \sum_{n \in \mathcal{N}_{a,k}} P_{a,k}^{n}},
\label{eqA52}
\end{align}
where $\mathcal{N}_{a,k}$ is the set of users subscribing to class $k$ in the coverage area of AP $a$. The utility function \(U(C^{n}_{a,k}(T_{j+1}))\) is approximated to simplify the optimization process. At each iteration $\epsilon$, it is linearized around the current transmit power $P_{a,k}^{n}(\epsilon)$ as:

\begin{align}
U(C^{n}_{a,k}(T_{j+1})) &\approx U(C^{n}_{a,k}(T_{j+1})) (\epsilon) + \nonumber \\
&\hspace*{-2cm}\quad \frac{dU}{dC} \cdot \frac{(R_{\text{PD}}H_{a}^n)^{2}}{\big (\xi^2 {\sigma}_{n}^2+ \sum_{\substack{a' \in A \\ a'\neq a}} P_{a',k}^{n}+ (R_{\text{PD}}H_{a}^n)^{2}P_{a,k}^{n} \big) \ln 2}  \cdot  \nonumber \\
&\hspace*{-2cm}\quad \Big(P_{a,k}^{n} -  P_{a,k}^{n}(\epsilon)\Big),
\label{eqA53}
\end{align}

\noindent where $\epsilon$ is the current iteration step. This approximation allows for convex reformulation of the original problem. At each iteration, the optimisation problem is reformulated as a convex problem:

\begin{align}
\max_{P_{\text{min}} \leq P_{a,k}^{n} \leq P_{\text{max}}} & \sum_{k \in \mathcal{K}} \sum_{n \in \mathcal{N}_{a,k}} \Bigg[ 
U(C^{n}_{a,k}(T_{j+1})) (\epsilon) + \nonumber \\
&\hspace*{-2cm}\quad \frac{dU}{dC} \cdot \frac{(R_{\text{PD}}H_{a}^n)^{2}}{\big (\xi^2 {\sigma}_{n}^2+ \sum_{\substack{a' \in A \\ a'\neq a}} P_{a',k}^{n} + (R_{\text{PD}}H_{a}^n)^{2}P_{a,k}^{n} \big) \ln 2}  \cdot  \nonumber \\
&\hspace*{-2cm}\quad \Big(P_{a,k}^{n} -  P_{a,k}^{n}(\epsilon)\Big) - EE~ P_{a,k}^{n} \Bigg].
\label{eqA54}
\end{align}
This problem is solved iteratively, subject to constraints (\ref{eq27}), (\ref{eq30}) and (\ref{eqA55}), which is

\begin{align}
    \varrho \leq P_{a,k}^{n} \leq P_{\text{max}},
\label{eqA55}
\end{align}

\noindent where

\begin{align}
    \varrho = \max \left( P_{min},\frac{(\xi^2 {\sigma}_{n}^2+ \sum_{\substack{a' \in A \\ a'\neq a}} P_{a',k}^{n})(2^{C_{\text{min},k}^{n} - 1})}{(R_{\text{PD}}H_{a}^n)^{2}} \right).
\end{align}
To solve the convex optimisation problem, we introduce a Lagrangian function. The Lagrangian function for the optimisation problem in \eqref{eqA54} under its constraints is: 

\begin{align}
   \mathcal{L}_a(P, \lambda, \mu) = \nonumber \\
   &\hspace*{-2cm} \sum_{k \in \mathcal{K}} \sum_{n \in \mathcal{N}_{a,k}} \Bigg(U(C^{n}_{a,k}(T_{j+1})) (\epsilon) + \nonumber \\
   &\hspace*{-2cm} \frac{dU}{dC} \cdot 
   \frac{(R_{\text{PD}}H_{a}^n)^{2}}{\big(\xi^2 {\sigma}_{n}^2+ \sum_{\substack{a' \in A \\ a'\neq a}} P_{a',k}^{n} + (R_{\text{PD}}H_{a}^n)^{2}P_{a,k}^{n}\big) \ln 2} \cdot \nonumber \\
   &\hspace*{-2cm} \big(P_{a,k}^{n} -  P_{a,k}^{n}(\epsilon)\big) - 
   EE ~ P_{a,k}^{n} 
   \Bigg) - \nonumber \\
   &\hspace*{-2cm} \Big(\sum_{k \in \mathcal{K}} \sum_{n \in \mathcal{N}_{a,k}} \lambda^{T_{j+1}}_{k,n}(\varrho - P_{a,k}^{n})\Big) -\nonumber \\
   &\hspace*{-2cm}\Big(\mu^{T_{j+1}}_{k,n} \Big( \sum_{k \in \mathcal{K}} \sum_{n \in \mathcal{N}_{a,k}}P_{a,k}^{n} - P_{\text{max}} \Big ) \Big).
\end{align}

\noindent where $\lambda^{T_{j+1}}_{k,n} \geq 0$ and $\mu^{T_{j+1}}_{k,n} \geq 0$ are the dual variables associated with the problem for the time period $T_{j+1}$, and these are fixed for $\tau$, the duration of each time period. As a result, the dual problem becomes the following:

\begin{equation}
 \min_{\lambda \geq 0, \mu \geq 0}\max_{ \varrho \leq P_{a,k}^{n} \leq P_{\text{max}}  } \mathcal{L}_a(P, \lambda, \mu), 
\end{equation}
with respect to $P_{a,k}^{n}$. The optimal power allocation \(P_{a,k}^{n}\) for fixed values of the
Lagrangian multipliers can be found through using the Karush-
Kuhn-Tucker (KKT) conditions by solving $\frac{d\mathcal{L}_a}{dP} = 0$.

\begin{align}
&\frac{d\mathcal{L}_a}{dP} = \frac{d}{dP} \bigg[U(C^{n}_{a,k}(T_{j+1})) (\epsilon) + \nonumber \\
&\hspace*{1cm} \frac{dU}{dC} \cdot \frac{(R_{\text{PD}}H_{a}^n)^{2}}{\big(\xi^2 {\sigma}_{n}^2+ \sum_{\substack{a' \in A \\ a'\neq a}} P_{a',k}^{n} + (R_{\text{PD}}H_{a}^n)^{2}P_{a,k}^{n}\big) \ln 2} \cdot \nonumber \\
&\hspace*{1cm} \big(P_{a,k}^{n} - P_{a,k}^{n}(\epsilon)\big)\bigg] -
   EE +\lambda^{T_{j+1}}_{k,n} + \mu^{T_{j+1}}_{k,n}.
\label{eqA56}
\end{align}

\noindent This yields

\begin{align}
&\frac{d}{dP} \bigg[U(C^{n}_{a,k}(T_{j+1})) (\epsilon) + \nonumber \\
&\hspace*{1cm} \frac{dU}{dC} \cdot \frac{(R_{\text{PD}}H_{a}^n)^{2}}{\big(\xi^2 {\sigma}_{n}^2+ \sum_{\substack{a' \in A \\ a'\neq a}} P_{a',k}^{n} + (R_{\text{PD}}H_{a}^n)^{2}P_{a,k}^{n}\big) \ln 2} \cdot \nonumber \\
&\hspace*{1cm} \big(P_{a,k}^{n} - P_{a,k}^{n}(\epsilon)\big)\bigg] = 
   EE - \lambda^{T_{j+1}}_{k,n} - \mu^{T_{j+1}}_{k,n}.
\end{align}
Now, $P_{a,k}^{n}$ can be allocated under $\varrho \leq P_{a,k}^{n} \leq P_{\text{max}}$ and a gradient descent method can be applied
in order to update the multipliers $\lambda^{T_{j+1}}_{k,n}$ and $\mu^{T_{j+1}}_{k,n}$ such that:

\begin{equation}
    \lambda^{T_{j+1}}_{k,n} ({\epsilon+1}) = \max \big( 0, \lambda^{T_{j+1}}_{k,n} ({\epsilon}) + \alpha_1 (\varrho -P_{a,k}^{n})\big), 
\end{equation}

\begin{equation}
\begin{aligned}
    \mu^{T_{j+1}}_{k,n}(\epsilon{+}1) = 
    \max \bigg(&0, \mu^{T_{j+1}}_{k,n}(\epsilon) \\
    &+ \alpha_2 \bigg(\sum_{k \in \mathcal{K}} \sum_{n \in \mathcal{N}_{a,k}} 
    P_{a,k}^{n} - P_{\text{max}}\bigg) \bigg)
\end{aligned}
\end{equation}

where $\alpha_1 \geq 0$, and $\alpha_2 \geq 0$. 

\subsection{Overall Algorithm}
{In real time, we employ our proposed PDP algorithm to estimate the number of active users subscribing to each class \(k\) and their distribution in the network, \( N_a^k(t^{o}_{j}) \), \( t^{o}_{j} \in \tilde{T}_j^k \), at time $t^{o}_{j}$ in the interval $T_j$. The estimates are used for predicting changes after \(\tau\) fixed duration in the next time period $T_{j+1}$. The predicted values are stored in \(\hat{N}^{a,k}_{(j+1)}\), and \( \tilde{N}_a^k(T_{j+1}) \) is calculated. After that, user association is carried, and the power is allocated. Note that, power allocation is performed using a Lagrangian-based optimisation procedure with iterative updates of the variables $\lambda^{T_{j+1}}_{k,n}$ and $\mu^{T_{j+1}}_{k,n}$, which remain fixed during the next interval $T_{j+1}$. This overall process yields an energy-efficient solution by proactively solving  the formulated optimization problem  in \eqref{eq24} at each interval \(\tau\) for real-time adaptation in dynamic OWC environments.}

\begin{table}[htbp]
\centering
\caption{Simulation Parameters}
\label{tab:simulation_parameters}
\begin{tabular}{|l|l|c|}
\hline
\textbf{Parameter}        & \textbf{Description}            & \textbf{Value}      \\ \hline
$h_{\text{Tx}}$           & Vertical separation             & 2 m                \\ \hline
$w_0$                     & Beam waist radius               & 5 $\mu$m           \\ \hline
$\lambda$                 & VCSEL wavelength                & 1550 nm            \\ \hline
$P_{i,v}$                 & VCSEL transmitted power                & 50 mW            \\ \hline
$L_c \times L_c$                 & VCSELs per optical AP                & 5 $\times$ 5 =25             \\ \hline
RIN                       & RIN Power Spectral Density      & $-155$ dB/Hz       \\ \hline
$n_{\text{lens}}$         & Lens refractive index           & 1.55               \\ \hline
$n_{\text{rec}}$          & Receiver refractive index            & 1.77               \\ \hline
$R_{\text{PD}}$           & Photodetector responsivity      & 0.7 A/W            \\ \hline
$N_{\text{PD}}$           & Number of photodetectors        & 5                 \\ \hline
FOV                       & ADR half-angle FOV              & $30^\circ$         \\ \hline
$F_n$                     & Preamplifier noise figure       & 5 dB               \\ \hline
$B$                       & VCSEL bandwidth                & 1.5 GHz              \\ \hline
BER                       & Pre-FEC BER                     & $10^{-3}$          \\ \hline
\end{tabular}
\end{table}
\section{Numerical Results and discussion}
\label{Numerical Results and discussion}
\subsection{System configuration}
In this section we present the numerical results for the VCSEL-based OWC indoor system described in Section \ref{system model}. We consider an indoor environment with dimensions of $\mathrm{5m \times 5m \times 3m}$, where 12 users with time-varying distributions in the room are served by 8 optical AP that are mounted on the ceiling and  uniformly distributed across the room to ensure full coverage. Each optical AP consists of a $L_{c} \times L_{c}$ micro lens VCSEL array, as described in \cite{10621108}. In this system, users are positioned 2 meters away from the VCSEL-based APs, and each user is equipped with an ADR consisting of 5 photodiodes, each photodiode points to a distinct direction to enhance the field of view. In the simulation, we model the user residence time within the designated service area to follow an exponential distribution with a mean duration as in \cite{main_PDP_paper}, and we set the prediction duration $\tau$ to 0.5, and 1 minute. Additionally, user mobility is modelled based on the parameters in \cite{PDP_paper_2,main_PDP_paper}. We use consumption factor [dB] as an energy efficiency metric as in \cite{Mingqing_EE}, and we consider 4-QAM and 16-QAM for the M-QAM simulation. We evaluate the performance of the proposed PDP-OPA algorithm and compare it to two other schemes; the first is a  distance-based user association with uniform power allocation scheme (Baseline), where power is allocated uniformly among users regardless of their subscriptions, i.e., traffic demand. The second is the PDP with uniform power allocation (PDP-UPA) scheme, in which AP-user association is first performed using the proposed PDP, followed by uniform power allocation based on these associations. The remaining simulation parameters are listed in Table I.
\begin{figure}[h]
\centering 
\includegraphics[width=\columnwidth]{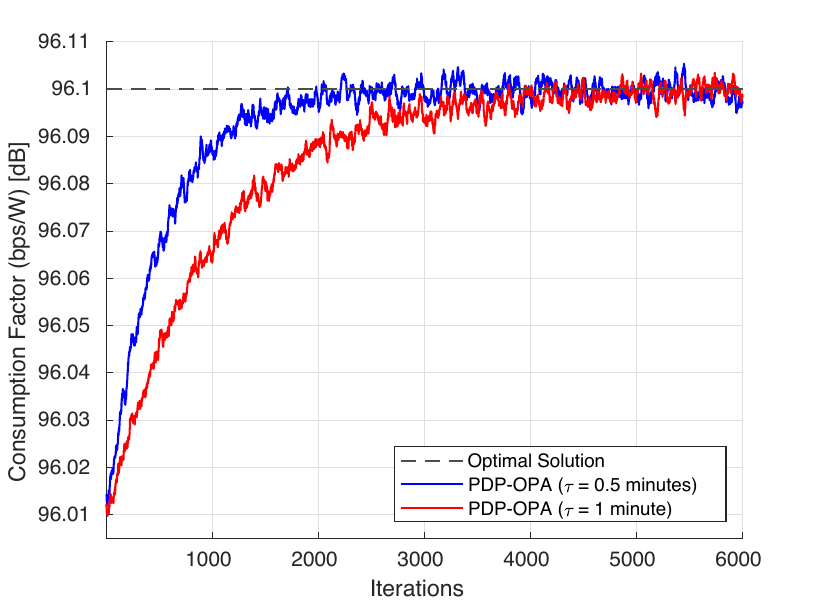}
\caption{PDP-OPA Convergence towards the optimal solution, $\mu=1.4$.} 
\label{fig:PDP Convergence} 
\end{figure}

\begin{figure}[h]
\centering 
\includegraphics[width=\columnwidth]{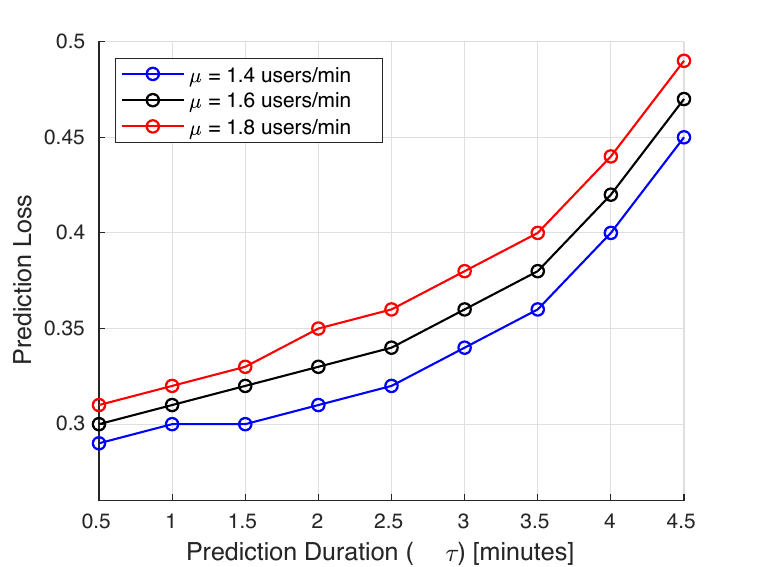}
\caption{Prediction loss versus prediction duration for different user arrival rates.}
\label{fig:Prediction Time vs Loss} 
\end{figure}
\subsection{Performance of the PDP-OPA Algorithm}

\begin{figure*}[t] 
  \centering
  \begin{subfigure}[t]{0.3\linewidth}
    \centering
    \includegraphics[width=\linewidth]{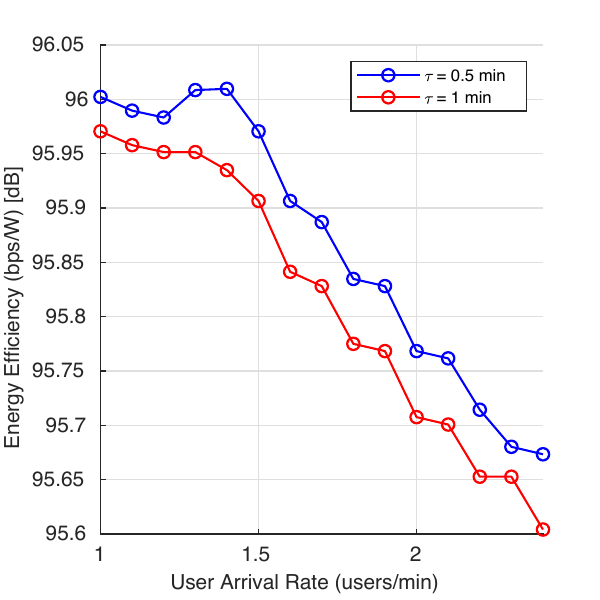}
    \caption{Video Streaming Users }
    \label{fig:sub1}
  \end{subfigure}\hfill
  \begin{subfigure}[t]{0.3\linewidth}
    \centering
    \includegraphics[width=\linewidth]{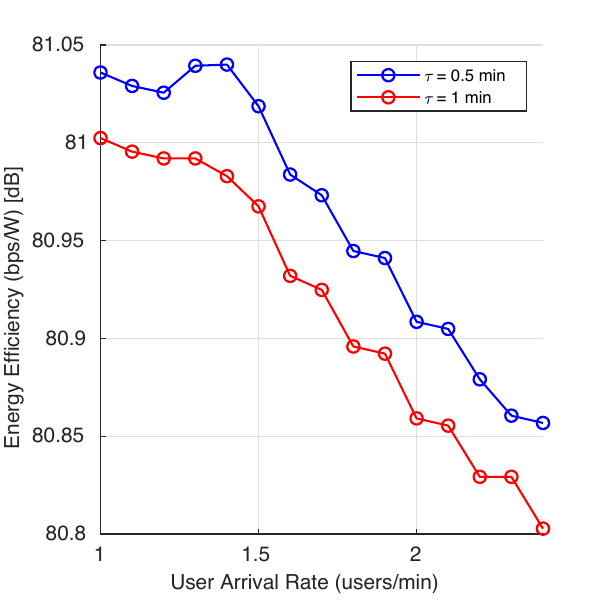}
    \caption{Web Users}
    \label{fig:sub2}
  \end{subfigure}\hfill
  \begin{subfigure}[t]{0.3\linewidth}
    \centering
    \includegraphics[width=\linewidth]{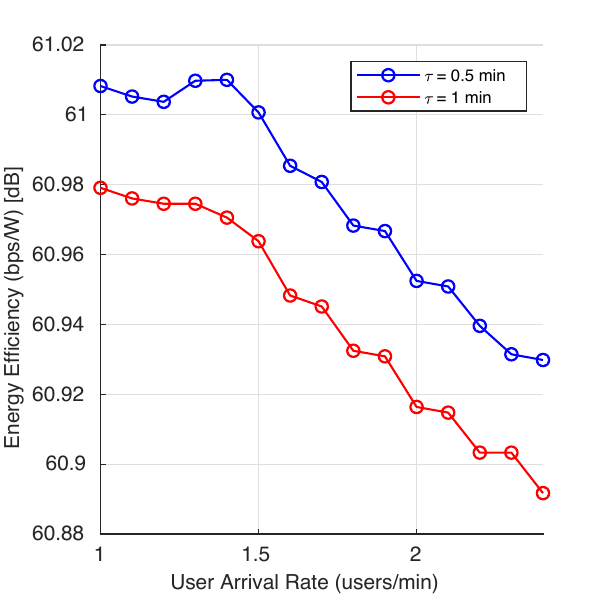}
    \caption{Voice Communication Users}
    \label{fig:sub3}
  \end{subfigure}
  
  \vspace{0.5cm}

  \begin{subfigure}[t]{0.3\linewidth}
    \centering
    \includegraphics[width=\linewidth]{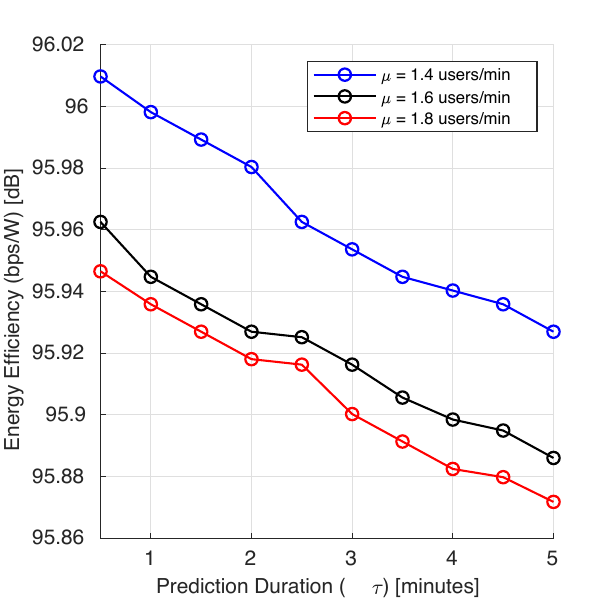}
    \caption{Video Streaming Users}
    \label{fig:sub4}
  \end{subfigure}\hfill
  \begin{subfigure}[t]{0.3\linewidth}
    \centering
    \includegraphics[width=\linewidth]{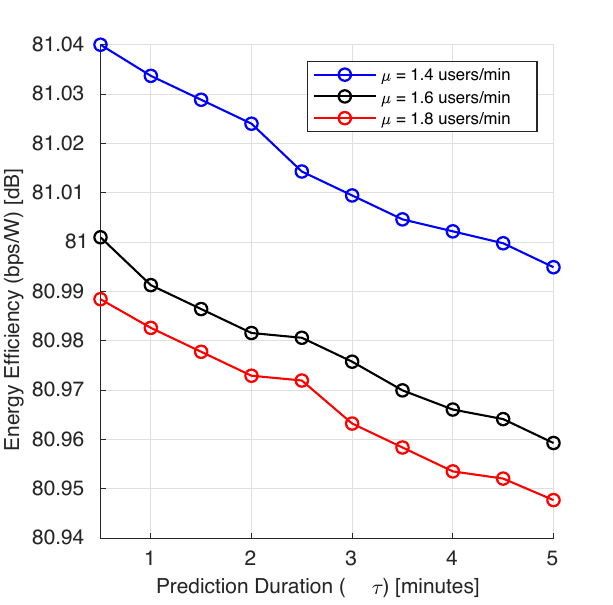}
    \caption{Web Users}
    \label{fig:sub5}
  \end{subfigure}\hfill
  \begin{subfigure}[t]{0.3\linewidth}
    \centering
    \includegraphics[width=\linewidth]{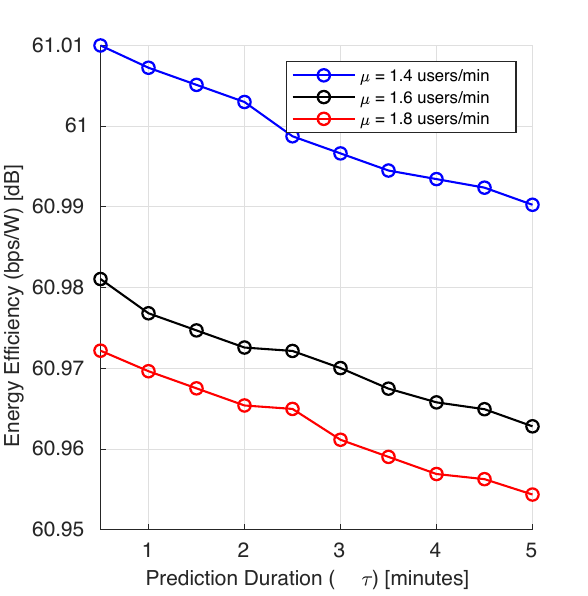}
    \caption{Voice Communication Users}
    \label{fig:sub6}
  \end{subfigure}
  
  \caption{Energy Efficiency for different classes with varying $\mu$ and $\tau$.}
  \label{fig:all_subfigures}
\end{figure*}

We examine the impact of prediction duration, \(\tau\), on the PDP-OPA algorithm’s performance. Figure \ref{fig:PDP Convergence} shows EE versus a set of iterations with user arrival rate  $\mu=1.4$. It can be seen that the PDP-OPA converges faster at \(\tau = 0.5\) than at \(\tau = 1\). This is due to the fact that  higher \(\tau\) reduces update frequency, delaying corrections in the predictions, and therefore, affecting the optimality of the proposed PDP-OPA. It should be noted that, while shorter \(\tau\) improves responsiveness and accuracy, it necessitate frequent estimations to maintain high accuracy in such  a rapidly changing environment. Conversely, longer \(\tau\) reduces adaptability, leading to higher losses, lower energy efficiency, and slower convergence. Therefore, an optimal \(\tau\) is needed to balance system reactivity with computational efficiency.

We further evaluate the impact of prediction duration \(\tau\) on the performance of the proposed PDP-OPA algorithm in terms of prediction loss  as shown in Figure \ref{fig:Prediction Time vs Loss}. It can be seen that  the prediction loss increases the prediction duration \(\tau\) due to less frequent updates, making the system slower to adapt to demand fluctuations, resulting in outdated predictions, higher variability and errors. {Moreover, from Figure \ref{fig:Prediction Time vs Loss}, it can be seen that prediction loss increases as the user arrival rate increases. This is due to the fact that as the number of users in the system increases, the system becomes more dynamic and complex, making it harder for the PDP to accurately make predictions. The increased variability and number of users in the system introduce more noise and uncertainty into the prediction process, reducing the model's accuracy and resulting in higher prediction loss.}
 
 In Figures \ref{fig:sub1}, \ref{fig:sub2} and  \ref{fig:sub3},  the effect of the user arrival rate $\mu$ on the energy efficiency per optical AP is depicted across different service classes $K$. The figures  show a decline in energy efficiency as the user arrival rate, i.e.,  the number of users, increases regardless of the user service class subscription. This is expected due to the fact that  higher arrival rates cause more competition for network resources, leading to inefficient allocations and reduced throughput. Additionally, higher arrival rates cause greater interference, further reducing system throughput and, consequently, the energy efficiency. In Figures \ref{fig:sub4}, \ref{fig:sub5}, and \ref{fig:sub6}, the energy efficiency per optical AP versus the prediction duration \(\tau\)  is depicted across different service classes $K$. It can be seen that   the energy efficiency across different service classes  decreases  as $\tau$ increases, since  less frequent updates lead to errors in predictions.  These errors impact the performance  of the proposed  PDP-OPA algorithm in determining  AP-user associations and power allocations, reducing system throughput and overall system efficiency. 

Interesting to note is that the subplots in Figure \ref{fig:all_subfigures} show that video streaming  exhibits the highest overall EE due to its high data throughput. It is also highly sensitive to changes in the network. It is worth mentioning that high-throughput environment demands stringent quality-of-service. Therefore, 
video streaming is extremely sensitive to changes: even small increases in prediction duration or user arrival rates can lead to significant efficiency losses. In other words,  providing video streaming service necessitates constant system adjustment to maintain high quality, otherwise, its efficiency drops steeply with increases in prediction duration and user arrival rate. In contrast, web applications show a more moderate decline. Voice communication is the least sensitive to changes due to its steady, predictable data transmission and lower bandwidth needs, which leads to a more stable performance, even as the load increases.

\subsection{Bit error rate and sum rate}
\begin{figure}[h]
\centering 
\includegraphics[width=\columnwidth]{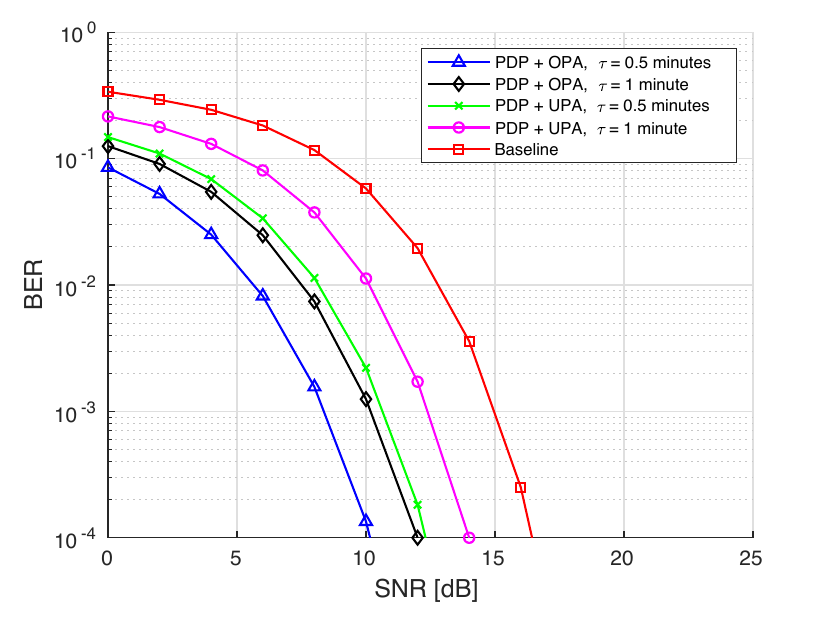}
\caption{BER of 4-QAM
DCO-OFDM for different scenarios.}
\label{fig:4QAM Simulation} 
\end{figure}

\begin{figure}[h]
\centering 
\includegraphics[width=\columnwidth]{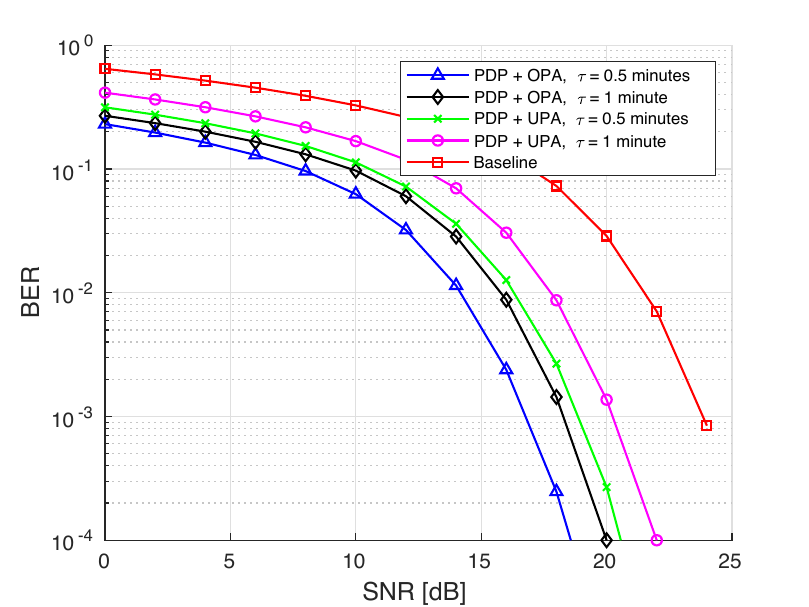}
\caption{BER of 16-QAM
DCO-OFDM for different scenarios.}
\label{fig:16QAM Simulation} 
\end{figure}

\begin{figure}[h]
\centering 
\includegraphics[width=\columnwidth]{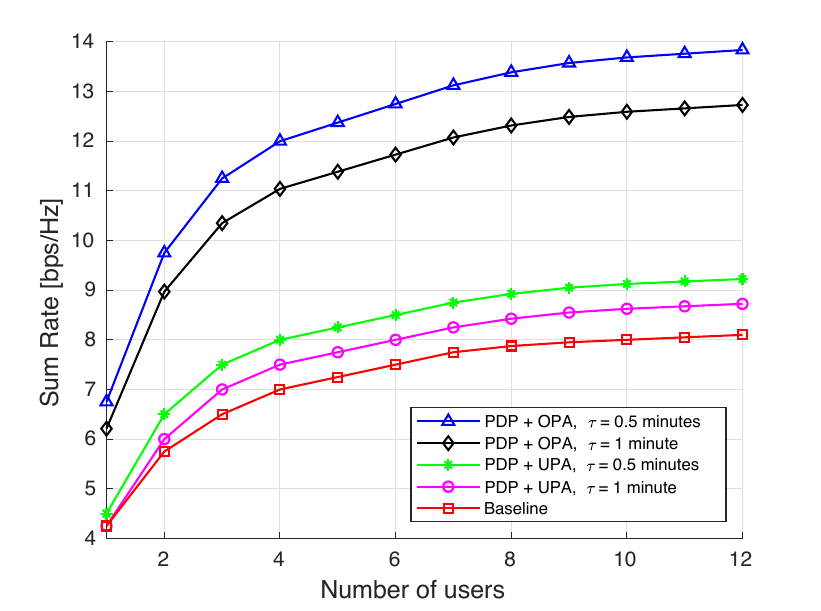}
\caption{Sum Rates versus the number of users in the network.}
\label{fig:EE per user} 
\end{figure}

\begin{figure}[h]
\centering 
\includegraphics[width=\columnwidth]{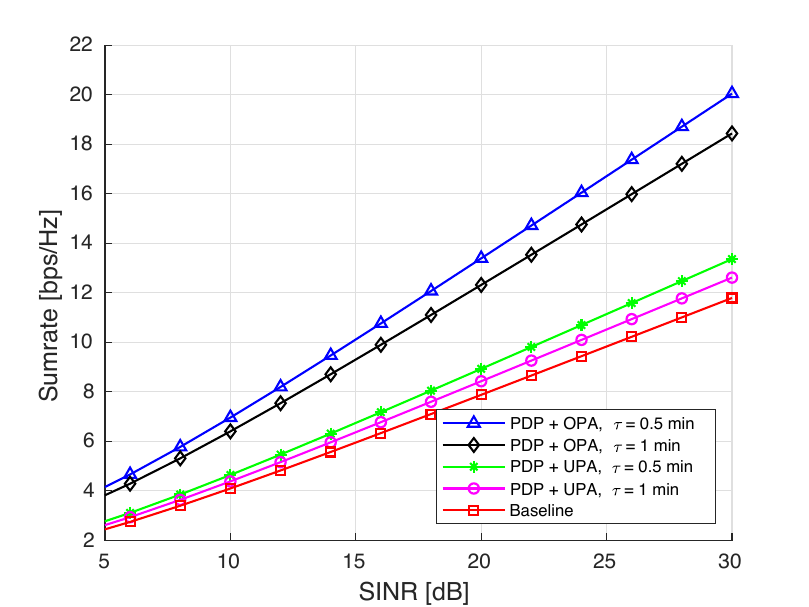}
\caption{Sum Rates versus the SNR of the network.}
\label{fig:EE vs SNR} 
\end{figure}

Figures \ref{fig:4QAM Simulation} and \ref{fig:16QAM Simulation} compare the BER performance of 4-QAM and 16-QAM for different simulation scenarios. From both the results, the implementation using 4-QAM produces smaller BER values relative to the 16-QAM. Also, the Baseline scheme is the least efficient, as it determines AP-user associations based on distance. Moreover,  it distributes power equally without considering user conditions. This results in some users  receiving more power than necessary, while other users struggle with insufficient power, leading to higher BER and requiring a higher SNR for improved performance. The PDP-UPA improves efficiency by optimising user association using the proposed PDP algorithm. However, its non-optimised power allocation still leads to inefficiencies. The PDP-OPA further enhances system performance by optimising both user association and power distribution, ensuring all users receive enough power to maintain a reasonable BER in relation to their subscriptions. This makes PDP-OPA the most efficient scheme, achieving the best BER performance at the lowest required SNR. Additionally, both PDP-OPA and PDP-UPA perform better at \(\tau = 0.5\) compared to \(\tau = 1\), highlighting the impact of prediction duration on system performance.

Figures \ref{fig:EE per user} and \ref{fig:EE vs SNR} show the sum rate against the number of users and SNR, respectively. The baseline scheme  is again the least efficient, as it fails to account for the network conditions and user demands, leading to lower overall throughput. The PDP-UPA improves performance by optimising user association using the proposed PDP, ensuring that users connect to the best optical APs. However, since power allocation remains uniform, capacity is not fully utilised, leading to inefficiencies. The PDP-OPA achieves the highest data rates due to its ability to adopt  dynamically to the network conditions. It adjusts user association and power allocation to ensure that all the users receive adequate data rate in relation to their traffic demands, where users subscribing  to video streaming receive higher resources compared to users subscribing to web and voice communication services.  As a result, the PDP-OPA outperforms all other schemes in both sum rate per user and sum rate per SNR, demonstrating the importance of our proposed strategy in optimising both power allocation and user association in discrete-time dynamic OWC networks.

\section{Conclusion}
\label{Conclusion}
In this paper, a predictive and proactive strategy was developed to enhance energy efficiency in discrete-time QAM-OFDM based OWC systems. We first modelled a VCSEL-based indoor OWC network that supports mobile users with time-varying user demands for heterogeneous services. The model incorporates ZF precoding for interference mitigation and classifies users into $K$ demand categories, enabling adaptive and efficient service-aware power allocation. To address the challenge of energy efficiency in such a dynamic environment, we formulated a discrete-time EE optimisation problem that maximises CF through joint user association and power allocation. Solving this problem in real time is computationally complex. To tackle this, we proposed a PDP-OPA framework. This approach proactively predicts traffic demand and user arrival and
departure to determine user association and dynamically
allocate power. Simulation results validated the effectiveness of the proposed PDP-OPA in providing solutions significantly close to the optimal one. Moreover, the proposed strategy achieved  improvements in  energy efficiency, sum rate, and BER compared to distance-based user association and uniform power allocation.

\bibliographystyle{IEEEtran}
\bibliography{Predictive_and_Proactive_Power_Allocation_For_Energy_Efficiency_in_Dynamic_OWC_networks}

\end{document}